\documentclass[draftclsnofoot,onecolumn,peerreview]{IEEEtran}

\usepackage{lineno}
\ifCLASSINFOpdf
   \usepackage[pdftex]{graphicx}
\else
   \usepackage[dvips]{graphicx}
\fi
\usepackage{amsmath}
\usepackage{subcaption}     

\newcommand{\iu}{\mathrm{i}}   
\newcommand{\D}{\mathrm{d}}
\newcommand{\aver}[1]{\left< #1 \right>}

\begin{document}
\linenumbers
\title{Full reconstruction of acoustic wavefields by means of pointwise measurements}

\author{Denis~V. Makarov,~
        Pavel~S. Petrov,% <-this % stops a space
\thanks{Department of Ocean Physics, POI FEB RAS, 43 Baltiyskaya Str., 690041, Vladivostok, Russia}% <-this % stops a space
\thanks{Manuscript received April 19, 2005; revised August 26, 2015.}}


\markboth{Journal of \LaTeX\ Class Files,~Vol.~14, No.~8, August~2015}%
{Shell \MakeLowercase{\textit{et al.}}: Bare Demo of IEEEtran.cls for IEEE Journals}



\maketitle

\begin{abstract}
Sound propagation in the ocean is considered.
We demonstrate a novel algorithm for full wavefield reconstruction
using pointwise measurements by means of a vertical array. 
The algorithm is based on the so-called discrete variable representation and
can be implemented both for tonal and pulse signals.
It is shown that the algorithm is robust against array distortions and ambient noise of moderate amplitude.
Efficiency of reconstruction is verified by means of numerical simulation 
with a model of a shallow-sea waveguide. It is found that the effect of bottom sound attenuation enables accurate reconstruction with arrays having relatively low density
of hydrophones.
\end{abstract}


% \fi
\IEEEpeerreviewmaketitle

\section{Introduction}

Efficient solving of various problems concerned with wave propagation often requires 
accurate reconstruction of a continuous wavefield relying upon data of pointwise measurements.
It is of the especial importance in inverse problems where even tiny features
of an interference pattern might give information about medium passed by a wave, or information about properties of a wave source.
In the context of underwater acoustics, it allows for straightforward extraction of wavefield constituents corresponding to individual modes \cite{Buck,WBP,AST,MPP,Labutina2016}, rays \cite{AET},
and the so-called stable components being Gaussian wavepackets with minimal uncertainty \cite{Artelny2018}.
%Such analysis of a wavefield
%can be useful for efficient solving of almost all practical and fundamental problems of underwater acoustics.
Expansion of a wavefield over individual modal contributions can be used, for example, for long-range environment reconstruction using mode-based schemes 
of hydroacoustical tomography \cite{Shang,Gorodetskaya,Jones,Radiophys}.
Also, detailed analysis of a wavefield structure is helpful for robust identification of propagation geometry, that is of great importance
for geoacoustic inversion \cite{Dumaz2019} and seismic surveys \cite{Rutenko2019,Abadi_Freneau}, 
as well as for various kinds of the matched field processing \cite{Lu_Yang_Duan,Viro-IEEE19,Liu}. 
Development of receiving-transmitting arrays allowing for full and accurate reconstruction 
is important in the context of applications associated with sound focusing, like 
underwater communication \cite{MPP,Volkov-vertical} and remote monitoring of the ocean bottom \cite{Lunkov2015,Lunkov-JASA2017}.
Detailed information about a wavefield structure facilitates experimental 
investigation of scattering physics \cite{Hege,Udo08,MorCol,ColosiMorozov,Colosi_Duda_Morozov,Udo-LOAPEX}, especially in the context of 
wave chaos \cite{Zas-UFN,Review03,RayWave,UFN,PRE87,JCA}.

Basically, the possibility of full reconstruction of a wavefield by means of a vertical array
follows from the Kotelnikov-Shannon-Nyquist theorem that determines the optimal sampling rate for representation of a continuous function by means of a discrete sequence. 
It implies exact reconstruction of a continuous wave (CW) field with wavelength $\lambda$ by means of a vertical array of hydrophones spaced by $\lambda/2$.
In the present study we show that optimal spacing can significantly exceed $\lambda/2$ due to the filtration of high-number modes.
The main result of the paper is the algorithm of exact reconstruction that takes into account boundary conditions in the waveguide. 
The algorithm is closely related to the so-called discrete variable representation.
Sensitivity of the algorithm to random fluctuations that can be caused by ambient noise or distortions of the array is also examined.

The paper is organized as follows.
The next section describes the discrete variable representation technique.
Reconstruction of tonal wavefields is considered in Section~\ref{CW}.
Short Section \ref{Choice} is devoted to array spacing values which provide accurate reconstruction.
Section \ref{Pulse} is devoted to reconstruction of broadband pulses.
In the Conclusion, we summarize and discuss the obtained results. Also we outline possible extensions of the algorithm and future work.

\begin{figure}[!ht]
\centering
\includegraphics[width=.70\textwidth]{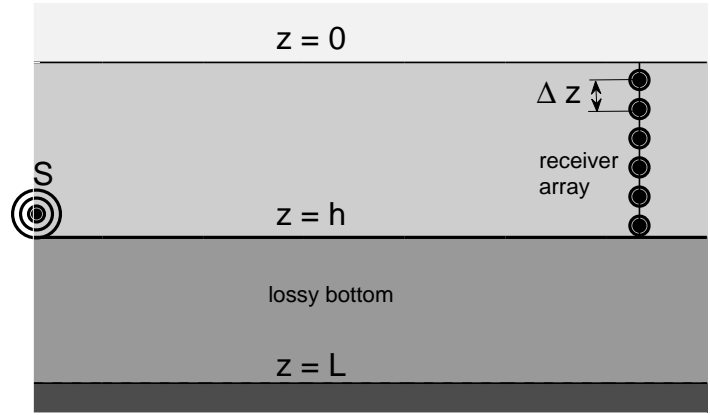}
\caption{A shallow-water waveguide with a sound source (S) and a vertical receiver array.}
\label{Fig-dom}
\end{figure}

\section{Discrete variable representation functions}
\label{SecDVR}
% PP: changed the sec title, DVR wasn't explained so far

The Kotelnikov-Shannon-Nyquist theorem considers
a continuous function $\Psi(\tau)$ having finite bandwidth $B$ (in Hz).
The theorem states that $\Psi(\tau)$
can be exactly reproduced by a discrete sequence of values 
$\Psi(\tau=nT_{\text{B}})$, where $T_{\text{B}}\le 1/(2B)$.
The reproduction can be accomplished by the Whittaker-Shannon interpolation formula
\begin{equation}
 \Psi(\tau) = \mu\sum\limits_{n=-\infty}^{\infty}\Psi(\tau_n)
 \chi_n(\tau).
 \label{WSh}
\end{equation}
where $\mu$ is some constant depending on the normalization condition adopted,
$\tau_n = nT_{\text{B}}$,
\begin{equation}
 \chi_n(\tau)=\mathrm{sinc}\left(\frac{\tau-\tau_n}{T_{\text{B}}}\right).
 \label{chi-sinc}
\end{equation}
Here the sinc function is defined as
$\mathrm{sinc}(x) = \sin{\pi x}/{\pi x}$.
The variable $\tau$ can be time or spatial coordinate.
The latter case anticipates that the Kotelnikov-Shannon-Nyquist theorem can be readily implemented in wave theory.

The present work is devoted to sound propagation in an underwater waveguide.
We are particularly interested in reproduction of an acoustic wavefield
relying upon measurements taken by a vertical array of receivers (see Fig.~\ref{Fig-dom}). 
Thus, we seek for the wavefield presentation of the following form
\begin{equation}
 \Psi(z) = \mu\sum\limits_{j=1}^{J}\Psi(z_j)
 \chi_j(z).
 \label{WSh-w}
\end{equation}
Here $z_j$ are depths of hydrophones being the array elements. 
As long as a waveguide is bounded in the vertical direction, the right-hand side of Eq.~(\ref{WSh-w}) is finite sum.

A wavefield must satisfy the boundary conditions in a waveguide, therefore, formula (\ref{chi-sinc}) for $\chi_j(z)$ is not valid
for guided wave propagation in general.
Acoustically soft ocean surface can be fairly modelled by the Dirichlet boundary condition
\begin{equation}
 \Psi(z=0)=0\,.
 \label{z0}
\end{equation}
Let us also assume that the bottom is flat and consists of two layers, a penetrable sediment and impenetrable basement.
At the sediment-to-basement boundary $z=L$ we impose Neumann boundary condition
\begin{equation}
 \frac{\D \Psi}{\D z}(z=L)=0.
 \label{zL}
\end{equation}
Our goal is to obtain such set of functions $\chi_j(z)$ that the wavefield reconstruction is accurately accomplished by the formula Eq.~\eqref{WSh-w}. It can be done using the so-called discrete variable representation theory \cite{Beck2000} (or shortly DVR), and the resulting set of functions $\chi_j(z)$ can be referred to as DVR functions.
DVR is extensively used in quantum mechanics in the context of multidimensional and/or many-body quantum calculations \cite{Beck2000,Pitsevich_Malevich}, 
construction of atomic Wannier states in periodic potentials \cite{Paul_Tiesinga}, and excitons in quantum dots \cite{Duron2017}. Basically, the main idea of the DVR approach is to construct a basis in which the position operator (in our case $z$-coordinate operator) is diagonalized. 

The first step of the derivation is the introduction of an auxiliary basis 
formed by spatial harmonics of the following form
\begin{equation}
 \phi_j = \sqrt{\frac{2}{L}}\sin{\frac{(2j-1)\pi z}{2L}},\quad 
 j = 1,2,\mathrm{...}\,.
 \label{phij}
\end{equation}
We see that this set satisfies boundary conditions \eqref{z0} and \eqref{zL}.
Function $\Psi$ can be expanded over harmonics (\ref{phij}),
%$\{\phi_j(z)\}_{j=1}^{\infty}$ 
%
\begin{displaymath}
    \Psi(z) = \sum_j a_j\phi_j(z).
\end{displaymath}
Amplitudes of wavefield expansion over functions $\phi_j$ are given by the formula
\begin{displaymath}
 a_j = \int\limits_{z=0}^L \phi_j(z)\Psi(z)\,\D z\,.
\end{displaymath}
We assume that a wavefield can be exactly reproduced using superposition of harmonics $\phi_j$ with $j\le j_{\max}$,  
i.~e. vertical wavenumber spectrum of $\Psi$ has finite bandwidth. 
%This condition is well satisfied for guided sound propagation 
%due to fast absorption of high-number modes.
Our assumption can be mathematically expressed as
\begin{equation}
 \sum\limits_{j=1}^{j_{\max}} |a_j|^2=\int\limits_{z=0}^L \Psi^*(z)\Psi(z)\,\D z.
 \label{constraint}
\end{equation}
The parameter $j_{\max}$ is finite for any wavefield except for the particular case $\Psi(z)=\delta(z-z_0)$,
where $\delta(z)$ is the delta function.
Using the auxiliary basis set, we can construct the $j_{\max}\times j_{\max}$ matrix
 $\mathbf{Z}$ with the entries
\begin{equation}
 Z_{mn} = \int\limits_{z=0}^{L}\phi_m^*(z) f(z) \phi_n(z)\,\D z.
 \label{Q}
\end{equation}
where $f(z)$ is a monotonic and invertible function.
Generally speaking, presentation of an acoustic wavefield 
in the form of expansion (\ref{WSh-w}) requires tridiagonal form 
of the matrix $\mathbf{Z}$ \cite{Beck2000}. It holds if
\begin{equation}
 f = \cos\frac{\pi z}{L}, \quad 0<z\le L.
 \label{fdef}
\end{equation}
One can find eigenvalues and eigenvectors of the matrix $\mathbf{Z}$
by solving the problem
\begin{equation}
    \mathbf{Z}\vec V_j = f_j \vec V_J, \quad
    j = 1, 2,\ldots, j_{\max}.
\end{equation}
The DVR theory allows one to find the DVR functions using eigenvectors $\vec V_j$ of the matrix $\mathbf{Z}$,
%
%\begin{equation}
% \mathbf{Z}\vec V_j = f_j\vec V_j,
%\end{equation}
%
%we can construct a new basis functions being superpositions of the auxiliary basis functions
%
\begin{equation}
\chi_{j}(z) = \sum\limits_{i=1}^{j_{\max}}V_{ij}\phi_{i}(z),
 \label{DVR}
\end{equation}
where $V_{ij}$ is the $i$-th entry of the $j$-th eigenvector $\vec V_j$. 
Formula (\ref{DVR}) ensures that DVR functions obey boundary conditions (\ref{z0}) and (\ref{zL}).
Eigenvectors can be found analytically:
\begin{equation}
 V_{ij}=\sqrt{\frac{2}{j_{\max}+1}}\sin\left[
 \frac{(i-\frac{1}{2})j\pi}{j_{\max}+1}
 \right]\,.
\end{equation}
Functions $\chi_j(z)$ also form complete orthogonal basis set if the condition (\ref{constraint}) is satisfied.
It means that an arbitrary wavefield satisfying (\ref{constraint}) can be expanded over DVR functions,
\begin{equation}
 \Psi(z) = \sum\limits_{j=1}^{j_{\max}} b_j\chi_j(z).
 \label{Psiz}
\end{equation}
The most important property of the DVR basis is the relation between anplitudes of this expansion
and local values of $\Psi(z)$ \cite{Beck2000},
\begin{equation}
 b_j = \sqrt{\Delta z}\Psi(z=z_j),
 \label{bj}
\end{equation}
where depth values $z_j$ are determined by eigenvalues of $\mathbf{Z}$,
\begin{equation}
 z_j = z(f_j).
\end{equation}
The property (\ref{bj}) provides means for reconstruction of a wavefield by means of pointwise measurements via a vertical array of hydrophones located
at depths $z_j$.

Eigenvalues $f_j$ are given by a simple formula 
\begin{equation}
 f_j = \cos\left(\frac{j\pi}{j_{\max} + 1/2}\right).
\label{fj} 
 \end{equation}
They determine depths of array elements
\begin{equation}
z_j = z(f_j)=j\Delta z,
\label{zj}
\end{equation}
where $\Delta z$ is array spacing,
\begin{equation}
 \Delta z = \frac{L}{j_{\max}+1}\,.
 \label{Dz}
\end{equation}
Thus, we see that the DVR representation using the auxiliary basis (\ref{phij})
corresponds to an equispaced vertical array. 
The array spacing is determined by the
parameter $j_{\max}$ that in turn is determined by the vertical Fourier spectrum of a wavefield.
Comparing (\ref{WSh-w}) with (\ref{Psiz}) and (\ref{bj}), we find that
the constant $\mu$ in the expansion (\ref{WSh-w})
is determined by the array spacing, 
\begin{equation}
\mu = \sqrt{\Delta z}.
\end{equation}
So, the formula (\ref{WSh-w}) becomes 
\begin{equation}
 \Psi(z) = \sqrt{\Delta z}\sum\limits_{j=1}^{J}\Psi(j\Delta z) \chi_j(z).
 \label{WSh-a}
\end{equation}
This formula allows one to reconstruct a continuous profile of an acoustic field by means of pointwise measurements using a vertical array,
if condition (\ref{constraint}) is satisfied. Accuracy of reconstruction depends on how exactly the condition (\ref{constraint}) is fulfilled.
It can be quantified as
\begin{equation}
 \epsilon=\left|\sum\limits_{j=1}^{j_{\max}} |a_j|^2 -\int\limits_{z=0}^L \Psi^*(z)\Psi(z)\,\D z\right|.
\end{equation}
%
%For $\epsilon=0$ the DVR functions form a complete orthonormal basis.
%
\begin{figure}[!ht]
\centering
  \includegraphics[width=.48\textwidth]{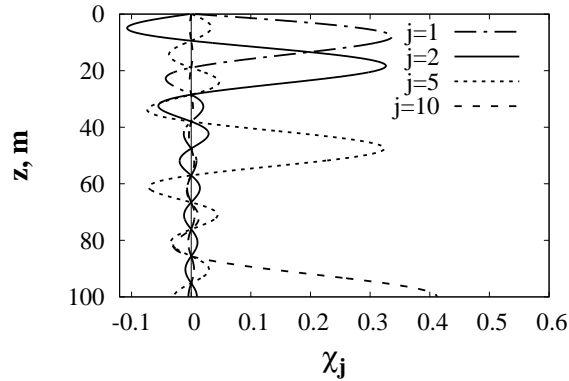}  
\caption{DVR functions corresponding to $j_{\max}=10$ for a waveguide with $L=100$~meters.
The corresponding depth eigenvalues are 9.52 meters for $j=1$, 19.04 meters for $j=2$, 47.62 meters for $j=5$, and 95.24 meters for $j=10$.
}
\label{Fig-chi}
\end{figure}

In Fig.~\ref{Fig-chi} several DVR functions corresponding to $j_{\max}=10$ are presented.
It is clear that each DVR function is localized in a vicinity of the corresponding depth eigenvalue. 
All DVR functions obey the boundary conditions given by Eqs.~\eqref{z0} and \eqref{zL}.

Number of hydrophones
needed for the full reconstruction of a vertical wavefield profile 
decreases with decreasing of $j_{\max}$.
On the other hand, limitations on a wavefield structure imposed by Eq.~\eqref{constraint}
also become more significant as $j_{\max}$ decreases. It means that there should be some optimal value of $j_{\max}$ that combines technical simplicity of the array with high quality 
of reconstruction.
%depending on acoustic wavelength.
As was mentioned before, in an unbounded isovelocity medium the optimal spacing value is given by the the Kotelnikov-Shannon-Nyquist theorem as
\begin{equation}
 \Delta z_{\infty} = \frac{\lambda}{2}.
 \label{Kotelnikov}
\end{equation}
However, typical underwater acoustical environment must be considered as a bounded vertically-stratified waveguide rather than an unbounded medium, since they are often
approximately homogeneous only in the horizontal directions. 
It is known that in such layered media field components with 
large vertical wavenumbers 
are strongly attenuated in course of sound propagation due to the interaction with the seabottom \cite{COA}. 
This effect can be also explained by ray escaping as considered in \cite{Chaos,Akust07}. 
It can be therefore anticipated that the optimal value of the array spacing can significantly exceed $\Delta z_{\infty}$.

It is worthwhile to mention that Eq.~\eqref{WSh-a} remains valid if we consider real-valued sound pressure field $u=\mathrm{Re}\Psi$ instead of 
a complex-valued wavefield $\Psi$. Then Eq.~(\ref{WSh-a}) reads
\begin{equation}
 u(z) = \mathrm{Re}\left(\sum\limits_{j=1}^{j_{\max}}b_j\chi_j(z)
 \right).
 \label{reconstruction}
\end{equation}
Note that strictly speaking the truncated representation of the wavefield Eq.~\eqref{WSh-w} (consisting of the first $J$ terms) is an approximation. Thus, the DVR functions constructed above in fact form a basis is a finite-dimensional subspace of the space of square-integrable functions on the interval $[0,L]$, and the functions $\chi_j(z)$ are associated with the restriction of the position operator to this subspace \cite{Beck2000}. As shown in the examples below, the inaccuracy arising from this simplification does not substantially affect the accuracy of the wavefield reconstruction. 

It might seem more natural to start the construction of the DVR functions with the basis $\phi_j(z)$ consisting of normal modes. However, it is essential for the DVR approach that matrix $\mathbf{Z}$ is tridiagonal, and it is unclear how to ensure this property for the normal modes in general case (i.e., how to choose the function $f(z)$ for a given sound speed profile and bottom properties in such a way that $\mathbf{Z}$ is tridiagonal).

\section{Reconstruction of time-harmonic wavefields}
\label{CW}

\subsection{Model of a waveguide}

\begin{figure}[!ht]
\centering
  \includegraphics[width=.98\textwidth]{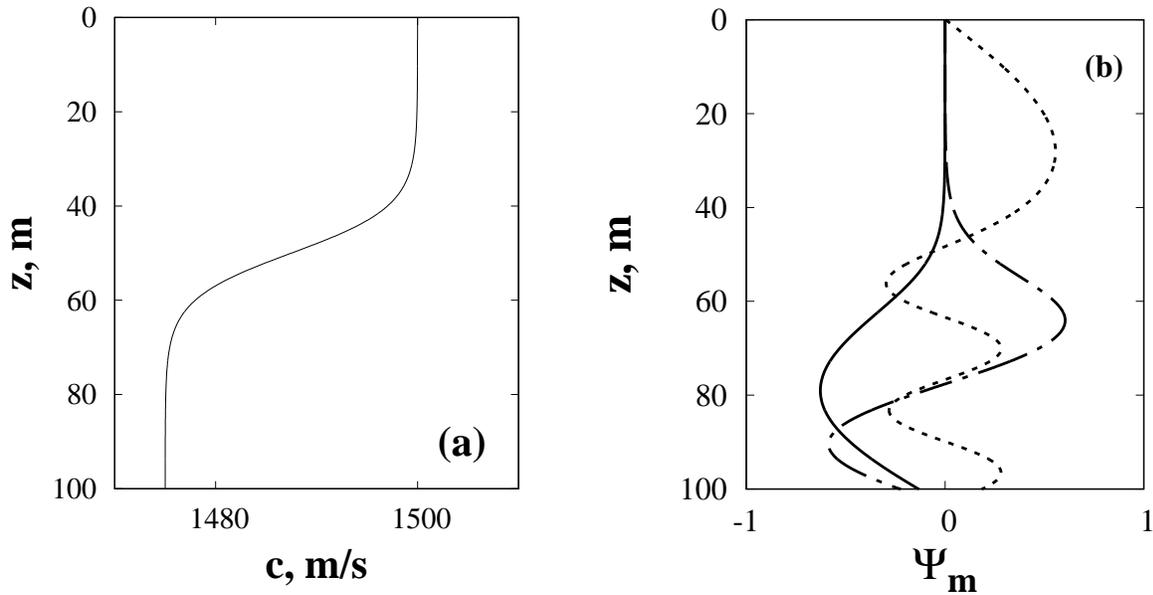}
\caption{(a) Sound speed profile. (b) Profiles of the first (solid line), second (dash-dotted line), and fifth (dashed line) normal modes for frequency 300 Hz.
}
\label{Fig-2}
\end{figure}

In this section we perform numerical validation of the accuracy of wavefield reconstruction using Eq.~\eqref{WSh-a} in the case of cw signals.
For numerical simulation, we use a model of a shallow-water waveguide with the sound-speed profile described by formula
\begin{equation}\label{cz}
c(z)=\left\{
  \begin{aligned}
    &c_{\mathrm{0}} - \frac{\Delta_{\mathrm{c}}}{2}\left[1 + \mathrm{tanh}\left(\frac{z-z_{\mathrm{c}}}{\Delta_{\mathrm{z}}}\right) \right],\quad 0\le z < h,\\
    &c_{\mathrm{b}},\quad h\le z\le L,
  \end{aligned}
 \right.
\end{equation}
where $c_{\mathrm{0}}=1500$ m/s is sound speed at the ocean surface, $\Delta_{\mathrm{c}}=25$ m/s, $z_{\mathrm{c}}=50$ m, $\Delta_{\mathrm{z}}=10$ m, $c_{\mathrm{b}}=1600$ m/s,
$L=300$~m.
The interface at the depth $z=h=100$~m separates the water and sediment layers. The profile is presented in Fig.~\ref{Fig-2}(a).
Waves propagating in this waveguide can be conditionally divided into two classes.
The first class corresponds to
the waves that propagate inside the near-bottom waveguide 
and do not reach the surface due to the refraction inside the water volume (more precisely, in the layer with sound speed gradient, or the thermocline).
The waves that propagate in the whole water volume and experience reflections from both natural boundaries,
i.e. from the bottom and the surface, belong to the second class. 
These two classes of waves are called reflected-refracted and reflected-reflected, respectively \cite{COA}.

Acoustical field in this shallow-water waveguide satisfies the 2D Helmholtz equation
\begin{equation}
    \frac{\partial^2 \Psi}{\partial r^2} + \frac{1}{r}\frac{\partial \Psi}{\partial r} + \rho\frac{\partial}{\partial z}\left[\rho\frac{\partial \Psi}{\partial  z}\right] + k_0^2n^2\Psi=0,
    \label{Helm}
\end{equation}
where $r$ is range, $\rho(z)$ is the density profile in the considered cross-section of the waveguide,
%$\rho$ is density,
$k_0=2\pi f/c_{\min}$,
is the reference wavenumber, $c_{\min}=c_{\mathrm{0}}-\Delta_{\mathrm{c}}$ is the minimal sound speed in a water column, and
$f$ is the sound frequency.
Refractive index $n(z)$ is given by formula
\begin{equation}
 n(z) = \frac{c_0}{c(z)} + 2\iu \alpha(f) \Theta(z-h),\quad
\end{equation}
where $\Theta(z-h)$ is the Heaviside function, $\alpha = 0.42\times10^{-6}f^2$ dB/m. 
The imaginary term describes sound attenuation in the sediment. 
The density profile is given by the step function
\begin{equation}
\rho(z)=\left\{
\begin{aligned}
&\rho_{\mathrm{wat}},\quad 0\le z \le h,\\
&\rho_{\mathrm{sed}},\quad h\le z \le L,
\end{aligned}
\right.
\label{rho}
\end{equation}
where $\rho_{\text{wat}}=1 \mbox{ g/cm}^3$,
$\rho_{\text{sed}}=1.7\mbox{ g/cm}^3$.
The sediment parameters correspond to conditions 
of the East China Sea \cite{Knobles-sandy09}.

Consider a wavefield emitted by a point source located at the depth $z=z_{\mathrm{s}}$. Invoking the far-field approximation, we can express the wavefield
as superposition of normal modes of the form
\begin{equation}
\begin{aligned}
 &\Psi(r,z) = \frac{\iu }{2\sqrt{2\pi r}}e^{-\iu \pi/4}\times\\
 &\times\sum\limits_{m=1}^M \frac{1}{\sqrt{k_{rm}}} \mathrm{e}^{\iu (k_{rm} + \iu \alpha_m)r}\Psi_m(z_{\mathrm{s}})\Psi_m(z),
 \end{aligned}
 \label{prz}
\end{equation}
where $\Psi_m(z)$ is the eigenfunction of the $m$-th normal mode, $M$ is number of normal modes belonging to the discrete spectrum (often called waterborne modes),
$k_{rm}$ and $\alpha_m$ are horizontal wavenumber and attenuation rate of the $m$-th normal mode, respectively, and ``c.~c.'' means complex conjugation.
Normal modes are solutions of the Sturm-Liouville problem \cite{COA}
\begin{equation}
\rho(z)\frac{\D}{\D z}\left[\frac{1}{\rho(z)}\frac{\D \Psi_m(z)}{\D z}\right]+
\left[k_0^2n^2(z) - k_{rm}^2\right]\Psi_m(z)=0,
 \label{StL}
\end{equation}
with the boundary conditions
\begin{equation}
 \Psi(0)=0,\quad \left.\frac{\D \Psi}{\D z}\right|_{z=L}=0\,
\end{equation}
and usual continuity conditions for the modal function \cite{COA}
\begin{equation}
\begin{aligned}
 \left.\Psi\right|_{z=L-0}&=\left.\Psi\right|_{z=L+0},\\ \frac{1}{\rho_{\mathrm{wat}}}\left.\frac{\D \Psi}{\D z}\right|_{z=L-0}&=\frac{1}{\rho_{\mathrm{sed}}}\left.\frac{\D \Psi}{\D z}\right|_{z=L+0}\,,
 \end{aligned}
\end{equation}
at the water-sediment interface $z=h$.
In the next paragraph we compare wavefields modelled using Eq.~(\ref{prz}) with their reconstructed counterparts. In all examples considered below Sturm-Liouville problem \eqref{StL} was solved using the code ac\_modes \cite{github_pavel} developed by the authors.

\subsection{Reconstruction in the noiseless environment}

Accuracy of reconstruction is quantified using the fidelity defined as \cite{Acoust17}
\begin{equation}
  F = \frac{1}{A_{\mathrm{exact}}A_{\text{est}}}\left|
 \int\limits_{z=0}^h \Psi_{\mathrm{exact}}^*(z)\Psi_{\mathrm{est}}(z)\,\D z
 \right|^2,
 \label{fidelity}
\end{equation}
where $\Psi_{\mathrm{exact}}(z)$ is a modelled wavefield considered as an ``exact'' solution, and $\Psi_{\mathrm{est}}(z)$ is a result of reconstruction using Eq.~(\ref{reconstruction}).
The normalization factors $A$ and $A_{\text{est}}$ are $L^2[0,h]$-norms of modelled and reconstructed wavefields, respectively,
\begin{displaymath}
 A_{\mathrm{exact}} = \int\limits_{z=0}^h |\Psi_{\mathrm{exact}}(z)|^2\,\D z,\quad
 A_{\text{est}} = \int\limits_{z=0}^h |\Psi_{\text{est}}(z)|^2\,\D z.
\end{displaymath}
According to the definition (\ref{fidelity}), $F=1$ if the modelled and reconstructed wavefields coincide, and tends to zero as differences between them grow.
The upper limit of integration in Eq.~(\ref{fidelity}) is taken of $h$, i.~e. the depth of water-sediment interface.
It means that we assume that a vertical array spans only the water layer, and the acoustical field inside the sediment is not taken
into account for the fidelity calculation. Then number of hydrophones in the array is given by the formula
\begin{equation}
    J = \text{floor}\left(j_{\max}\frac{h}{L}\right),
\end{equation}
where $\text{floor}(x)$ is the function that produces rounding down of $x$ if $x$ is not integer.

%
%\begin{figure}[!ht]
%\centering
%\includegraphics[width=.48\textwidth]{Fig3.eps}  
%\caption{Fidelity as the function of the signal frequency for the array consisting of 10 hydrophones. 
%The dashed line indicates the upper boundary of the confidence range (CR) in the frequency domain. 
%Distance from the source: (a) 1 km, (b) 10 km, (c) 40 km.}
%\label{Fig-pure10}
%\end{figure}
%

\begin{figure}[!ht]
\centering
\begin{subfigure}[b]{0.32\columnwidth}
\includegraphics[width=\columnwidth]{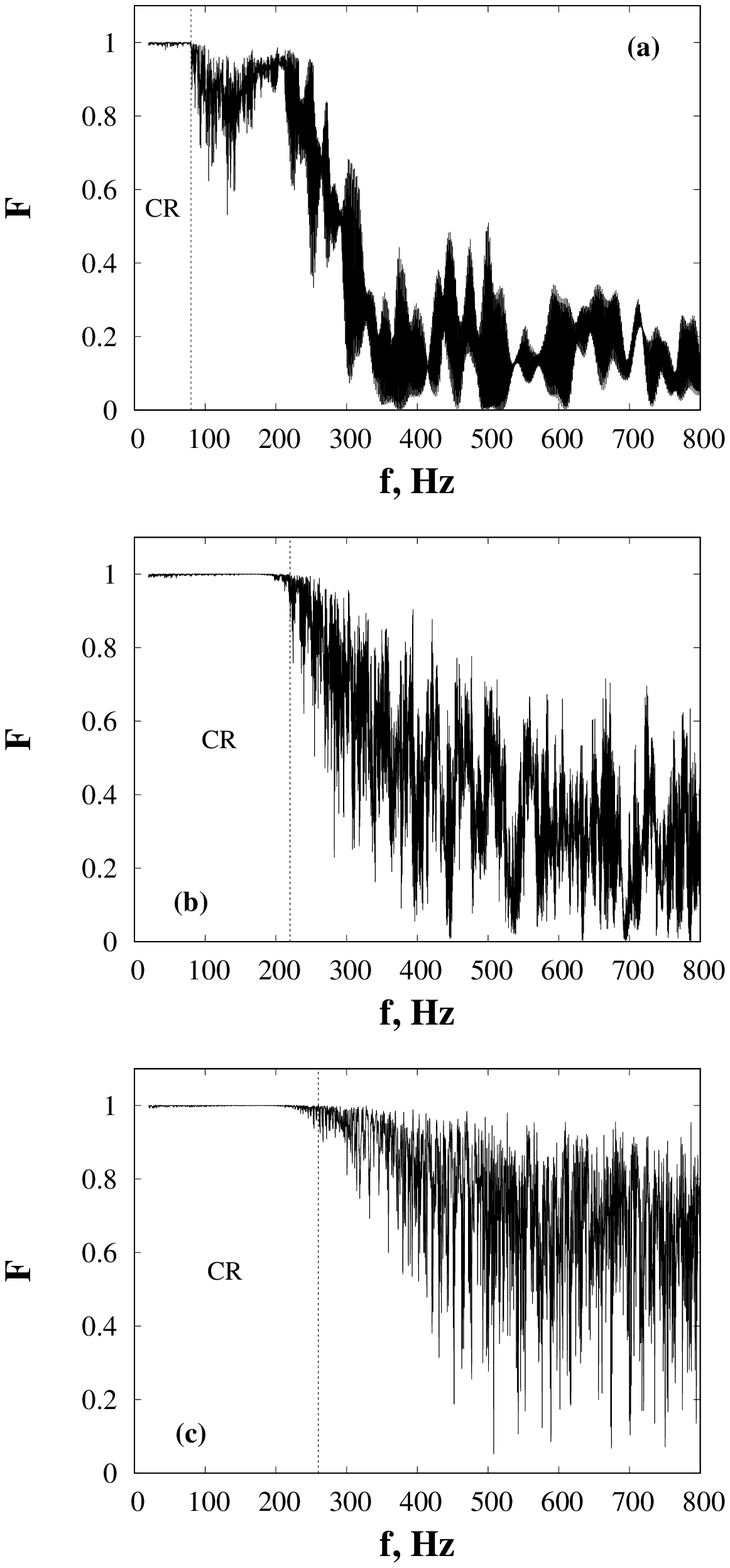} 
%\caption{}
\end{subfigure}
\begin{subfigure}[b]{0.32\columnwidth}
\includegraphics[width=\columnwidth]{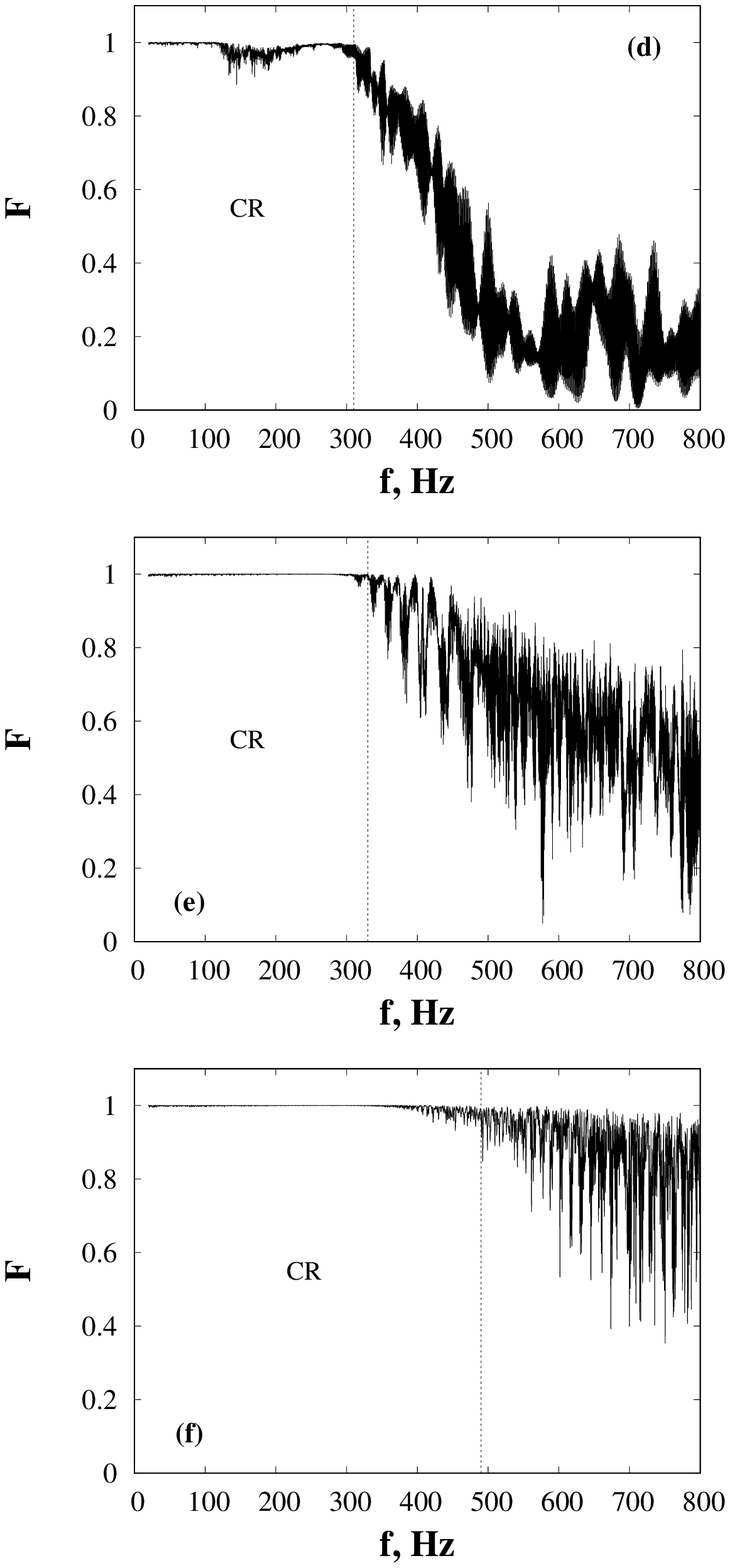} 
%\caption{}
\end{subfigure}
\begin{subfigure}[b]{0.32\columnwidth}
\includegraphics[width=\columnwidth]{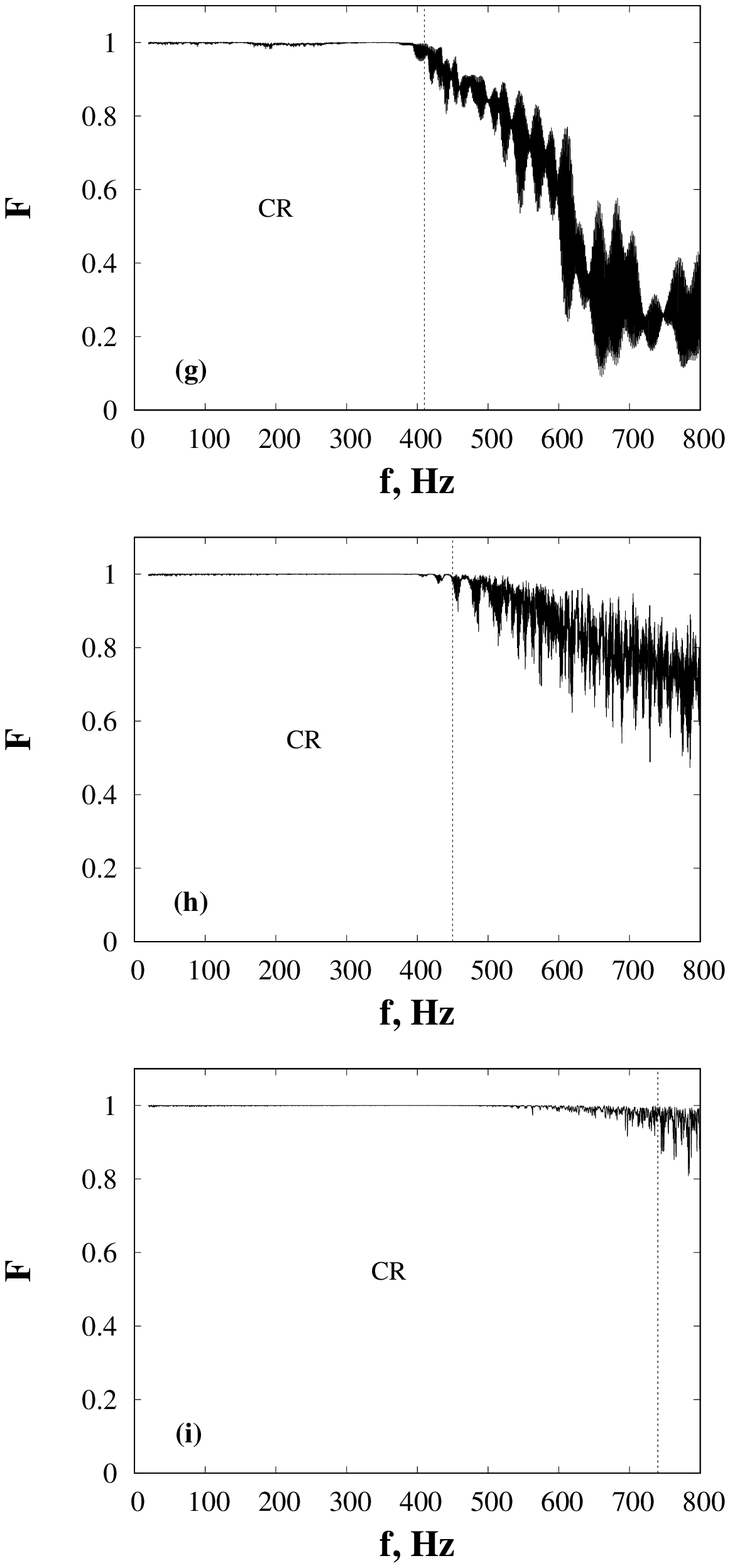} 
%\caption{}
\end{subfigure}
\caption{Fidelity as the function of the signal frequency. The dashed line indicates the upper boundary of the confidence range (CR) in the frequency domain. Number of hydrophones in the array:   10 for panels (a)-(c), 15 for panels (d)-(f), and 20 for panels (g)-(i). Distance from the source:
1 km for the upper row of panels, 10 km for the middle row, and 40 km for the lower row.} 
\label{Fig-combined}
\end{figure}

In the present section we consider only cw wavefields. Reconstruction of broadband pulses will be considered in Sec.~\ref{Pulse}.
We consider wavefields created by a point source located near the bottom, at $z=99$~m, at various ranges.
One can conditionally define confidence range of an array in the frequency space
as a sufficiently broad frequency interval where 
\begin{equation}
 F>0.9.
 \label{F09}
\end{equation}
Figs.~\ref{Fig-combined}(a)-(c) demonstrate the dependence of fidelity on the signal frequency for the array consisting of 10 hydrophones, i.~e. for $j_{\max}=30$.
The corresponding receiver spacing in the array is 9.52 m. 
We see that width of the frequency confidence range strongly depends on distance.
In the case of $r=1$~km, the array provides accurate reconstruction of a wavefield for frequencies up to
$80$~Hz. It is very close to the critical frequency predicted by the Kotelnikov-Shannon-Nyquist theorem that is equal to 75 Hz.
Increasing of distance results in the suppression of of high-number modes due to the bottom attenuation, and therefore, scale of vertical interference pattern
significantly increases.
It remarkably expands the confidence range: its upper boundary
is 220 Hz for $r=10$~km, and 260 Hz for $r=40$~km.

%
%\begin{figure}[!ht]
%\centering
%\includegraphics[width=.48\textwidth]{Fig4.eps}
%\caption{The same as in Fig.~\ref{Fig-pure10}, but for the array of 15 hydrophones.}
%\label{Fig-pure15}
%\end{figure}
%

%
%\begin{figure}[!ht]
%\centering
%\includegraphics[width=.48\textwidth]{Fig5.eps} 
%\caption{The same as in Figs.~\ref{Fig-pure10} and \ref{Fig-pure15}, but for the array of 20 hydrophones.}
%\label{Fig-pure20}
%\end{figure}
%

Decreasing of array spacing enhances its capability to resolve fine-scale features of the interference pattern and
leads to significant broadening of the confidence range 
(see Figs.~\ref{Fig-combined}(d)-(f) and \ref{Fig-combined}(g)-(i)).
In the case of the array with 15 hydrophones, the upper boundary of the confidence range
is 330 Hz for $r=10$~km, and 490 Hz for $r=40$~km.
Data for $r=1$~km presented in Fig.~\ref{Fig-combined}(d)
deserves particular attention: frequency dependence of fidelity is surprisingly non-monotonic, 
and there is an interval of lowered fidelity in the middle of 
the confidence range, near $f\simeq 145$~Hz. 
This kind of non-monotonicity is associated with the competition of two factors:
decreasing of wavelength that results in the reconstruction quality deterioration, and growth of
sound attenuation that facilitates suppression of high-number modes.
The latter factor tends to restore fidelity with increasing frequency, as shown in Fig.~\ref{Fig-combined}(d).

\begin{figure}[!ht]
\centering
\includegraphics[width=.48\textwidth]{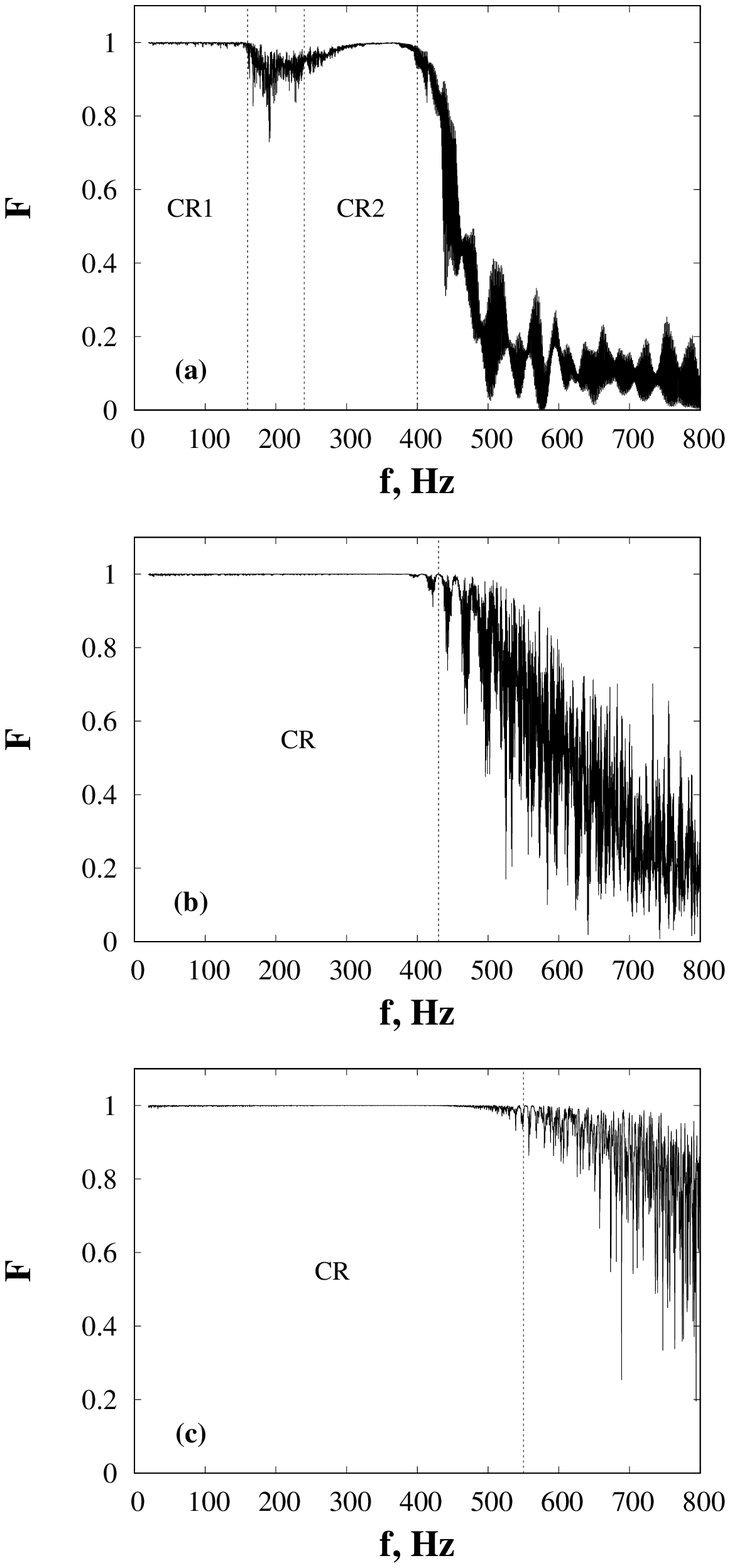}
\caption{The same as in Fig.~\ref{Fig-combined}(g)-(i), but for a source located at $z=1$ m. Two parts of the splitted confidence range in the case of $r=1$~km 
are denoted as ``CR1'' and ``CR2''.}
\label{Fig-pure_up}
\end{figure}

Striclty speaking, the data presented in Fig.~\ref{Fig-combined}(d) exhibits pointwise violation of inequality (\ref{F09}) at 
$f=144.4$ Hz, where $F=0.88$, but we consider it as non-essential.
A more pronounced example of the fidelity non-monotonicity in the low-frequency range 
is presented in Fig.~\ref{Fig-pure_up}(a), where lowering of fidelity leads to
splitting of the confidence range into two parts.

%that is proportional to $f^2$ and implies more extensive damping of high-number modes.
The array with 20 hydrophones provides broad confidence range for all considered distances.
The upper boundary is 410 Hz for $r=1$~km, 450 Hz for $r=10$~km, and 740 Hz for $r=40$~km.

Distribution of acoustical energy over modes depends on the source position. Placing the source closer to the ocean surface, we can enhance 
impact of high-number modes and reduce their filtering. It worsens quality of reconstruction and diminishes fidelity values.
Figure \ref{Fig-pure_up} represents fidelity dependence on frequency for the array with 20 hydrophones and the point source
located near the surface, at $z=1$ m. 
We see that width of the frequency confidence range is diminished as compared with data presented in Fig.~\ref{Fig-combined}(g)-(i).
In addition, the confidence range in the case of $r=1$~km is divided into two parts. It is a manifestation of the aforementioned effect 
of fidelity non-monotonicity associated with enhanced mode filtration for high frequencies.

\subsection{Impact of noise and fluctuations}

\begin{figure*}[!ht]
\centering
\includegraphics[width=.98\textwidth]{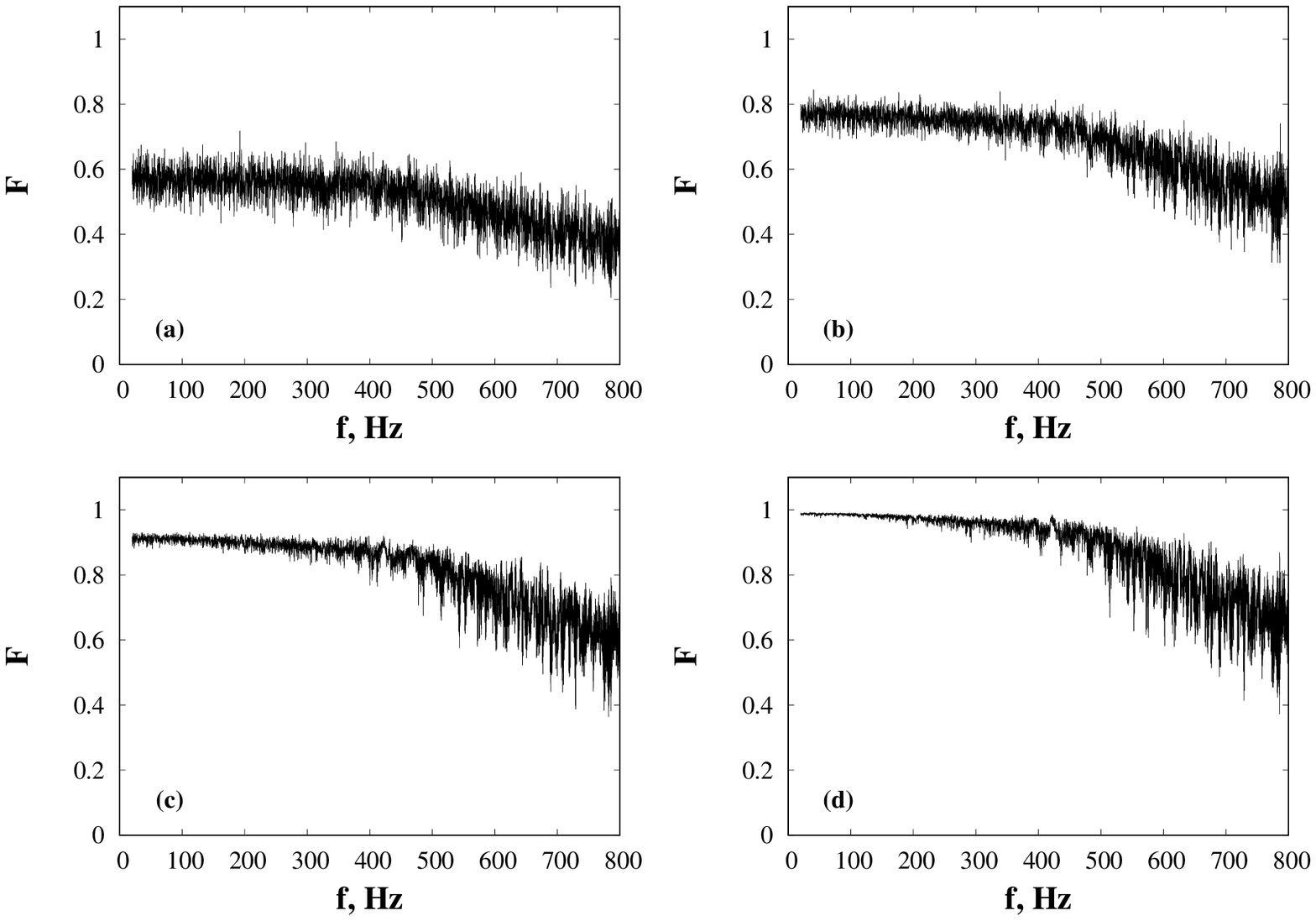}  
\caption{Frequency dependence of fidelity averaged over 10 realizations of noise for the array with 20 hydrophones. (a) $\mathrm{SNR}_{\mathrm{dB}}=1$ dB, 
(b) $\mathrm{SNR}_{\mathrm{dB}}=5$ dB, (c) $\mathrm{SNR}_{\mathrm{dB}}=10$ dB, (d) $\mathrm{SNR}_{\mathrm{dB}}=20$ dB.}
\label{Fig-SNR1}
\end{figure*}

It is reasonable to examine sensitivity of the reconstruction to ambient noise and inaccuracies in hydrophone positioning.
Effect of noise can be modelled as random perturbation $\xi_j$ of measured values of acoustic pressure,
\begin{equation}
\Psi_{\text{measured}}(z_j) = \Psi(z_j) + \xi_j.
\label{um1}
\end{equation}
We considered several models of ambient noise. The most significant destructive effect onto accuracy of reconstruction comes from 
spatially white noise,
when values of $\xi_j$ with different $j$ are uncorrelated, and their variances don't depend on $j$.
Models of noise generated by the ocean surface affect reconstruction to smaller extent.
As long as we are interested in testing reliability of reconstruction under the noise disturbances, 
it is reasonable to consider the most worse case. Therefore, we present below the results obtained with spatially white noise.
Taking into account that number of statistically independent transmissions might be limited in an actual experiment,
we conduct statistical analysis with relatively poor ensemble of ten realizations.

\begin{figure*}[!ht]
\centering
\includegraphics[width=.98\textwidth]{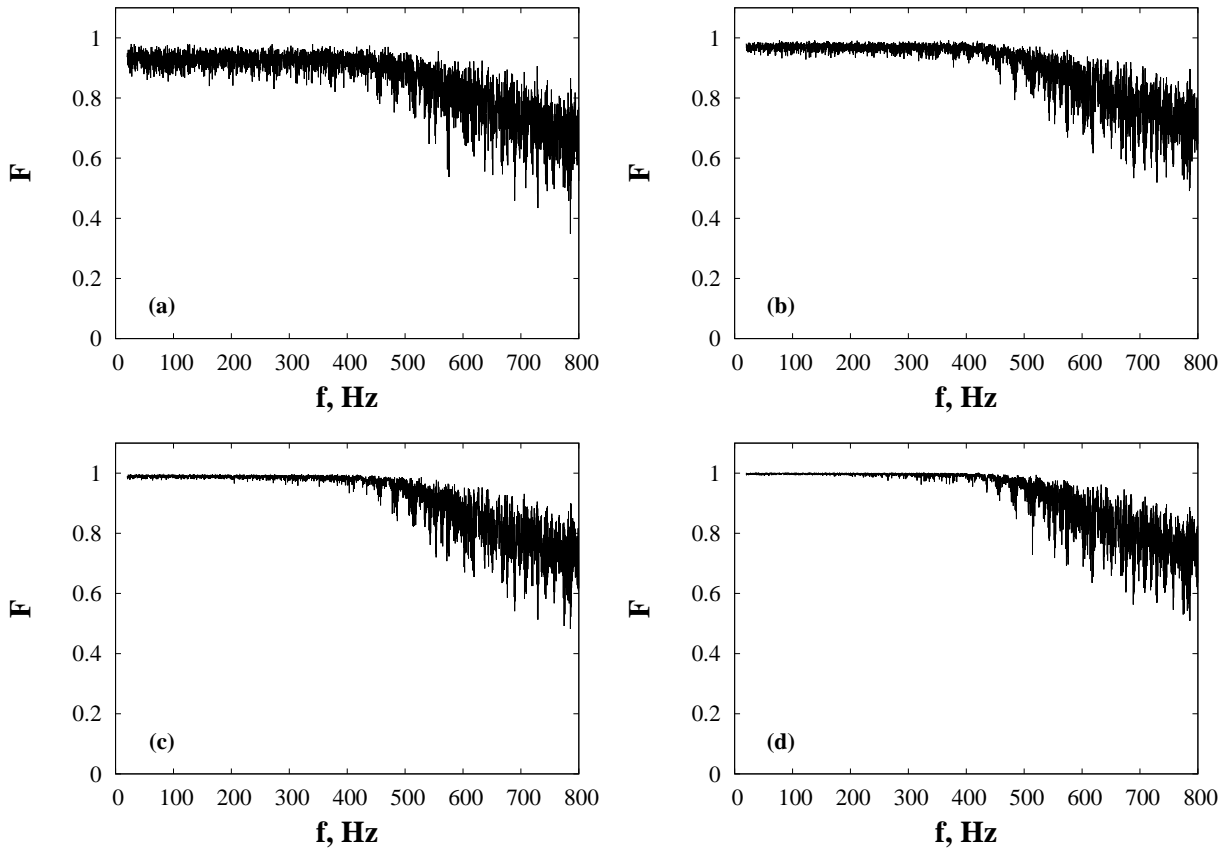}  
\caption{Frequency dependence of fidelity calculated with the averaged data.
The array consists of 20 hydrophones. (a) $\mathrm{SNR}_{\mathrm{dB}}=1$ dB, 
(b) $\mathrm{SNR}_{\mathrm{dB}}=5$ dB, (c) $\mathrm{SNR}_{\mathrm{dB}}=10$ dB, (d) $\mathrm{SNR}_{\mathrm{dB}}=20$ dB.}
\label{Fig-SNR2}
\end{figure*}

Reconstruction errors caused by ambient noise can be amplified by displacements of the array hydrophones from the depths determined by Eq.~(\ref{zj}).
Perturbation of similar kind was considered in \cite{Eliseevnin2002}.
Horizontal displacements can be reasonably considered as small compared with horizontal sound wavelength, therefore, their effect is negligible
for the array performance.
When constructing a model of such displacements, we can assume that the upper and lower ends of the array are tightly fixed at the surface and bottom, respectively.
Then, the displacement field can be presented as sum of sine modes with random amplitudes. We restrict ourselves to the case of two modes, 
and the displacement profile is written as
\begin{equation}
 \zeta(z) = \frac{\varsigma}{\sqrt{2}}\left(\zeta_1\sin\frac{\pi z}{h} + \zeta_2\sin\frac{2\pi z}{h} \right),
 \label{eta}
\end{equation}
where $\nu_1$ and $\nu_2$ are statistically independent Gaussian random variables with zero mean and unit variance.
In computations, r.m.s. amplitude of displacement $\varsigma$ is taken 1 m. 
The resulting perturbation is a combination of ambient noise and random hydrophone displacements, therefore Eq.~\eqref{um1} has the form
\begin{equation}
\Psi_{\text{measured}}(z_j) = \Psi(z_j+\zeta) + \xi_j,
\label{um2}
\end{equation}
Figure \ref{Fig-SNR1} demonstrates frequency dependence of fidelity for various values of signal-to-noise ratio (SNR).
SNR is evaluated as 
\begin{equation}
 \mathrm{SNR}_{\mathrm{dB}} = 10\,\mathrm{log}_{10}\frac{\left( \sum_{j=1}^J |\Psi_{\mathrm{exact}}(z_j)|^2  \right)}{\sum_{j=1}^J |\xi_j|^2},
\end{equation}
where function $\Psi_{\mathrm{exact}}(z)$ is computed by Eq.~(\ref{prz}).
The data corresponds to the array with $J=20$ hydrophones and a point source located at $z=99$~m. 
Note that we simulate relatively difficult measurement conditions with low SNR values. It is found that
the effect of displacements on the fidelity is weak as compared to that of the noise. Not surprisingly, the noise reduces accuracy of the reconstruction. 
Satisfactory level of the fidelity can be obtained with SNR of 10 dB or higher. 
However, influence of the noise can be significantly reduced by averaging 
measured values $\tilde \Psi$ over realizations before inserting them into Eq.~(\eqref{WSh-a}),
\begin{equation}
 \aver{\Psi}(z_j) = \frac{1}{N}\sum\limits_{n=1}^N\tilde \Psi(z_j)\,.
 \end{equation}
Here $N$ is number of realizations. Averaging substantially improves
the quality of reconstruction, yielding accurate reconstruction even in the case of $\text{SNR}=1$~dB (see Fig.~\ref{Fig-SNR2}).

\begin{figure}[!ht]
\centering
\includegraphics[width=.48\textwidth]{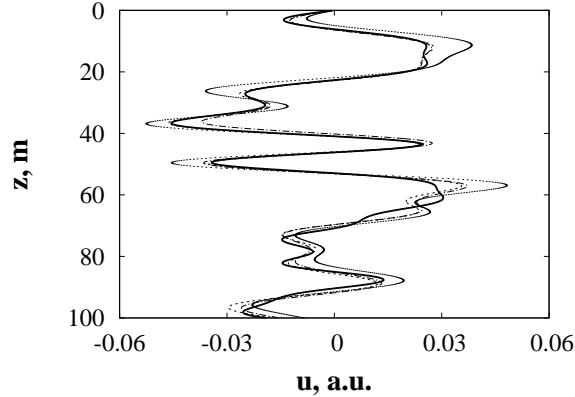}  
\caption{Comparison of the original wavefield profile (bold solid line) with reconstructed ones. 
Dashed curve corresponds to reconstruction in the absence of fluctuations, fidelity $F=96.8$\%. Dotted line corresponds to reconstruction
for an individual realization of ambient noise with SNR 10 dB and array displacements with $\varsigma=1$~m, $F=85.4$\%. 
Dashed dotted line corresponds to reconstruction in the presence of fluctuations, with the preliminary averaging of a wavefield over 
10 realizations, $F=94.9$\%.
The data corresponds to the array with 20 hydrophones. Signal frequency is 500 Hz. Propagation range is 10 km.
}
\label{Fig-wsampl}
\end{figure}

Quality of reconstruction could be estimated by means of data presented in Figure \ref{Fig-wsampl}.
It illustrates the original and reconstructed wavefield profiles in the same plot (the signal frequency in this case is 500 Hz). 
In this particular example,
the fidelity value In the absence of fluctuations is 96.8\%. 
We see that the profiles are very close to each other, excepting the vicinity of $z=60$~m, where the reconstruction 
inaccurately reproduces form of an interference peak. 

%After removing effect of noise and array displacements, we see that discrepancies
% in the lower half of a waveguide are stronger. 
%Indeed,  contribution of the short-wavelength
%component of vertical wavefield spectrum is greater in the vicinity of the sound-speed minimum, that is located near the bottom.
%Apparently associated with impact high-number modes.

%

\section{Selection of the spacing between hydrophones}
\label{Choice}

As it follows from Eq.~\eqref{Dz}, the array spacing $\Delta z$ is determined by the total number of DVR functions $j_{\max}$.
For fixed depth of the sediment-basement interface $L$, its value can have a discrete set of values.
In practice, however, accurate information about value $L$ can be lacking. 
Furthermore, the low-order modes propagating predominantly inside the water column are almost insensitive to value of $L$, if the width of the sediment layer is sufficiently large.
Consequently, if we are interested in sound propagation over distances of few kilometers, or more, when the wavefield is mainly formed by low-order modes, we have certain freedom in choosing $L$.
It means that there are no strict limitations on the value of spacing $\Delta z$.
For arbitrarily chosen value of $\Delta z$, we can replace the ``actual'' value of $L$ by a fictitious one $L'$
that is determined using the formula
\begin{equation}
 L' = (j_{\max}+1)\Delta z.
\end{equation}
It leads to a slight modification of basis functions \eqref{phij},
\begin{equation}
 \phi_j = \sqrt{\frac{2}{L'}}\sin{\frac{(2j-1)\pi z}{2L'}},\quad 
 j = 1,2,\dots \,.
 \label{phij2}
\end{equation}
It should be noted that the value of $j_{\max}$ has to be sufficiently large in order to eliminate the effect 
of the sediment-basement interface onto structure of DVR functions inside the water column.

Thus, it turns out that the algorithm of wavefield reconstruction using the basis of DVR functions can be utilized
for any equally spaced vertical array, provided that the spacing is sufficient to reproduce vertical spectrum of a wavefield.
This property will be used in the next section for studying reconstruction of wavefields corresponding to broadband signals.

\section{Reconstruction of pulse wavefields}
\label{Pulse}

\begin{figure}[!ht]
\centering
   \includegraphics[width=.48\textwidth]{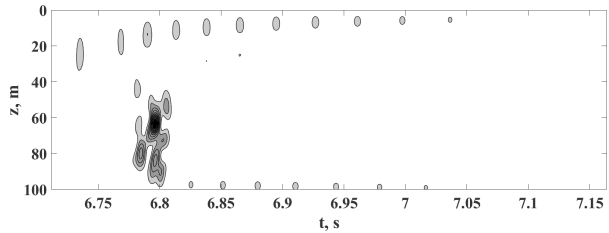}  
\caption{A wavefield in the time-depth plane at range $10$~km,
the case of a broadband pulse with center frequency $f_{\text{c}}=240$ Hz.
The pulse is created by a point source located near the bottom, at $z=99$~m.
}
\label{Fig-pulse}
\end{figure}

Transient wavefield $\tilde u$ corresponding to a sound pulse can be synthesized from cw wavefields $u$ using the formula \cite{COA}
\begin{equation}
 \tilde \Psi(t,z) = \int s(\Omega)\Psi(r,z,\Omega)\mathrm{e}^{-\iu \Omega t}\,\D \Omega,
\end{equation}
where $\Omega=2\pi f$ is the angular frequency, and $s(\Omega)$ is the spectrum of the signal emitted by the source. In our examples we consider broadband signals with the spectrum
\begin{equation}
 s(\Omega) = \frac{T}{\sqrt{2\pi}}\exp\left[
 -\frac{(\Omega-\Omega_{\mathrm{c}})^2T^2}{2}
 \right]\,,\quad 
 T = \frac{\sqrt{2\pi}}{\Delta\Omega}\,,
\end{equation}
where $\Delta\Omega = \Omega_{\text{c}}/2$.
Arrival pattern of such signal in the time-depth plane for the pulse with center frequency 240 Hz is demonstrated in Fig.~\ref{Fig-pulse}.
%The center frequencies used for simulation are 120, 240 and 480 Hz.

Fidelity for a pulse wavefield reconstruction can be defined in the same spirit as for cw wavefields, i.~e.
\begin{equation}
 F = \frac{1}{\tilde A_{\mathrm{exact}}\tilde A_{\text{est}}} \int\limits_{t=0}^{\infty}
 \int\limits_{z=0}^h  \tilde \Psi_{\mathrm{exact}}^*(t,z)\tilde \Psi_{\mathrm{est}}(t,z) 
 \,\D t\,\D z
\end{equation}
where the normalization factors are given by formulae
\begin{displaymath}
\begin{aligned}
\tilde A_{\mathrm{exact}} &= \int\limits_{t=0}^{\infty} \int\limits_{z=0}^h |\tilde \Psi_{\mathrm{exact}}(t,z)|^2\,\D t\,\D z,\\
 \tilde A_{\text{est}} &= \int\limits_{t=0}^{\infty} \int\limits_{z=0}^h |\tilde \Psi_{\text{est}}(t,z)|^2\,\D t\,\D z.
\end{aligned}
 \end{displaymath}%
\begin{figure}[!ht]
\centering
   \includegraphics[width=.48\textwidth]{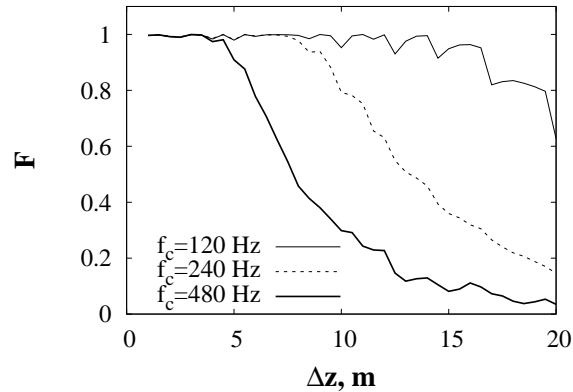}  
\caption{Fidelity of a pulse wavefield vs array spacing for broadband pulses with different center frequencies.
}
\label{Fig-fp}
\end{figure}

Figure \ref{Fig-fp} shows fidelity decay with increasing of the array spacing $\Delta z$
for pulses with center frequencies of 120, 240 and 420 Hz.
The pulses are emitted by the point source located near the bottom at $z=99$~m.
As one would expect, the confidence range in the $\Delta z$ space is inversely proportional
to the pulse center frequency.
Fidelity is nearly one within some limited interval of $\Delta z$ that can be considered as
the confidence range, and rapidly decreases outside it.
Arrays with $\Delta z\le 4.5$ m provide almost exact reconstruction of all considered pulses.

\section{Conclusion}
In the present study we demonstrate a novel algorithm for the wavefield reconstruction relying upon the data collected by a vertical array of equally spaced hydrophones. 
The proposed algorithm can be considered a generalization of a Whittaker-Shannon interpolation procedure
onto the case of underwater acoustic waveguides.
Validity of the algorithm requires boundedness of wavefield Fourier spectrum.
We see that sound attenuation in the bottom basically improves quality of reconstruction due to filtering of high-number modes. It implies that the fidelity dependencies presented in this paper are not universal and depend on the properties of a particular waveguide under consideration, as well as
on the form of the transmitted signal. 

The key idea of the algorithm is representation of a wavefield as expansion over
DVR functions. The coefficients of the expansion are equal to acoustic wavefield values at an equispaced grid. It means that they can be measured using an equispaced vertical array, where each pointwise measurement contributes to the coefficient of only one DVR function in the expansion. It is very convenient 
from the viewpoint of error elimination: inaccuracies of data measured by different hydrophones
do not interfere and can be processed separately. In the present paper we show that simple
averaging over data from few transmissions is sufficient for substantial reduction of 
noise-induced inaccuracy. Such averaging can be very effective if noise correlation time
is much smaller than the timescale of waveguide variability.

Another important advantage of the presented approach is the absence of explicit dependence
of DVR functions on signal frequency. This feature makes is very well-suited for reconstruction of broadband pulses. Also, DVR functions do not  depend on hydrological properties
of a waveguide. It means that this approach can be readily implemented for analysis of data
obtained by stationary  systems of acoustic monitoring of marine environment variability.

Numerical simulations in this paper are conducted with a range-independent model of a waveguide. It would be interesting to examine effect of horizontal inhomogeneities onto accuracy of the proposed algorithm. For instance, it is reasonable to expect that sound scattering by random inhomogeneities should lead to additional pumping of high-number modes thereby reducing effect of their filtration.
Another issue that should be addressed in future work is generalization of the algorithm to the case of arrays with variable spacing between receivers. In particular, in many cases it can be useful to decrease hydrophone spacing near the waveguide axis, where impact of shorter vertical wavelengths is more pronounced. Definitely, the feasibility of the wavefield reconstruction in a deep ocean propagation scenario also deserves investigation.

% or


\section*{Acknowledgment}

This work was carried out in the framework of the POI FEB RAS Program ``Mathematical simulation and analysis of dynamical processes in the ocean'' (No.~0271-2019-0001).

\ifCLASSOPTIONcaptionsoff
  \newpage
\fi




%\bibliographystyle{IEEEtran}
%\bibliography{biblio}


\begin{IEEEbiographynophoto}{Denis Makarov}
D. Makarov graduated from Krasnoyarsk State University in 1999 (diploma in physics). 
After the university he joined V.I.Il'ichev Pacific Oceanological Institute (POI),
the Laboratory of Nonlinear Dynamical Systems, as a Ph.D. student. 
He received Ph.D. degree in acoustics in 2004 and became a senior researcher of POI.
In 2015, he received Dr.Sci. degree in theoretical physics and was promoted to leading research associate of POI.
Denis Makarov won several awards and prizes, including medal of the Russian Academy of Sciences for young scientists,
L.M. Brekhovskikh medal from the Russian Acoustical Society, V.I.Il'ichev and U.H.Kopvillem awards from the Far-Eastern Branch of the Russian Academy of Sciences,
and stipend of the Dynasty Foundation.
The main fields of his research are underwater acoustics, chaos theory, quantum optics, and ultracold atoms.

\end{IEEEbiographynophoto}

\begin{IEEEbiographynophoto}{Pavel Petrov}
P. Petrov graduated from Irkutsk State University in 2006 (diploma in pure mathematics with distinction). 
In the same year he joined V.I.Il'ichev Pacific Oceanological Institute (POI) as a Ph.D. student. 
He received his Ph.D. degree in theoretical physics in 2010 and continued his work at POI as a senior researcher. 
In 2019 he became head of laboratory of Geophysical Hydrodynamics. 
In parallel, P. Petrov also taught mathematics 
(algebra, group theory, computer algebra, calculus) at Far Eastern Federal University starting as teaching assistant (2007) 
and later being promoted to assistant professor (2009) and associate professor (2010).

P.~Petrov had several scholarships at leading Russian research institutions, 
including St. Petersburg Department of Steklov Mathematical Institute and Institute for Problems in Mechanics (Moscow). 
He also had a postdoc scholarship in acoustics at University of Haifa (Israel), 
three scholarships in applied mathematics at University of Wuppertal (Germany) and a scholarship at University of Lorraine (France).

His research interests include mathematical physics, computational underwater acoustics, geoacoustic inversion, semi-analytical methods for 3D sound propagation.
  
\end{IEEEbiographynophoto}


\end{document}